\documentclass[aps,prl,twocolumn,showpacs,superscriptaddress,amsmath]{revtex4-1}

\pdfoutput=1

\usepackage{graphicx}
\usepackage{xspace}
\usepackage{hyperref} 
\usepackage[usenames]{color} 
\usepackage{ulem}
\usepackage{multirow}
\usepackage{amsmath}
\usepackage{accents}

\begin{document}



\title{
First combined analysis of neutrino and antineutrino oscillations at T2K
}

\newcommand{\superk}           {Super-Kamiokande\xspace}       
\newcommand{\nue}                {$\nu_{e}$\xspace}
\newcommand{\nuebar}           {$\bar{\nu}_{e}$\xspace}
\newcommand{\numubar}        {$\bar{\nu}_{\mu}$\xspace}
\newcommand{\numu}             {$\nu_{\mu}$\xspace}
\newcommand{\nutau}             {$\nu_{\tau}$\xspace}
\newcommand{\nux}                {$\nu_{x}$\xspace}
\newcommand{\numunue}       {$\nu_{\mu} \rightarrow \nu_{e}$\xspace}
\newcommand{\numunux}       {$\nu_{\mu} \rightarrow \nu_{x}$\xspace}
\newcommand{\numunutau}    {$\nu_{\mu} \rightarrow \nu_{\tau}$\xspace}

\newcommand{\tonethree}       {$\theta_{13}$\xspace}
\newcommand{\tonetwo}         {$\theta_{12}$\xspace}
\newcommand{\ttwothree}       {$\theta_{23}$\xspace}
\newcommand{\ssttmue}          {$\sin^2 2 \theta_{{\mu}e}$\xspace}
\newcommand{\sstonethree}    {$\sin^2 2 \theta_{13}$\xspace}
\newcommand{\ssttwothree}    {$\sin^2 2 \theta_{23}$\xspace}

\newcommand{\msqonetwo}   {$\Delta m^2_{12}$\xspace}
\newcommand{\msqonethree} {$\Delta m^2_{13}$\xspace}
\newcommand{\msqtwothree} {$\Delta m^2_{23}$\xspace}
\newcommand{\absmsqtwothree} {$|\Delta m^2_{23}|$\xspace}
\newcommand{\msqmue}        {$\Delta m^2_{{\mu}e}$\xspace}
\newcommand{\msqmumu}     {$\Delta m^2_{\mu\mu}$\xspace}

\newcommand{\enu}               {$E_{\nu}$\xspace}
\newcommand{\pmu}              {$p_{\mu}$\xspace}
\newcommand{\amome}         {$E_{e}$\xspace}
\newcommand{\evis}              {$E_{vis}$\xspace}
\newcommand{\pizero}           {$\pi^{0}$\xspace}
\newcommand{\pizerogg}       {$\pi^{0}\to\gamma\gamma$\xspace}

\newcommand{\degree}        {$^\circ$\xspace}
\newcommand{\evsq}          {\mathrm{eV^2}}
\newcommand{\evmsq}          {\mathrm{eV^-2}}

\newcommand\brabar{\raisebox{-4.0pt}{\scalebox{.25}{
\textbf{(}}}\raisebox{-4.0pt}{{\_}}\raisebox{-4.0pt}{\scalebox{.25}{\textbf{)
}}}}
\newcommand{\bbar}[1]{\accentset{\brabar}{#1}}
\newcommand{\bplus}[1]{\accentset{(+)}{#1}}


\newcommand{\INSTEE}{\affiliation{University of Bern, Albert Einstein Center for Fundamental Physics, Laboratory for High Energy Physics (LHEP), Bern, Switzerland}}
\newcommand{\INSTFE}{\affiliation{Boston University, Department of Physics, Boston, Massachusetts, U.S.A.}}
\newcommand{\INSTD}{\affiliation{University of British Columbia, Department of Physics and Astronomy, Vancouver, British Columbia, Canada}}
\newcommand{\INSTGA}{\affiliation{University of California, Irvine, Department of Physics and Astronomy, Irvine, California, U.S.A.}}
\newcommand{\INSTI}{\affiliation{IRFU, CEA Saclay, Gif-sur-Yvette, France}}
\newcommand{\INSTGB}{\affiliation{University of Colorado at Boulder, Department of Physics, Boulder, Colorado, U.S.A.}}
\newcommand{\INSTFG}{\affiliation{Colorado State University, Department of Physics, Fort Collins, Colorado, U.S.A.}}
\newcommand{\INSTFH}{\affiliation{Duke University, Department of Physics, Durham, North Carolina, U.S.A.}}
\newcommand{\INSTBA}{\affiliation{Ecole Polytechnique, IN2P3-CNRS, Laboratoire Leprince-Ringuet, Palaiseau, France }}
\newcommand{\INSTEF}{\affiliation{ETH Zurich, Institute for Particle Physics, Zurich, Switzerland}}
\newcommand{\INSTEG}{\affiliation{University of Geneva, Section de Physique, DPNC, Geneva, Switzerland}}
\newcommand{\INSTDG}{\affiliation{H. Niewodniczanski Institute of Nuclear Physics PAN, Cracow, Poland}}
\newcommand{\INSTCB}{\affiliation{High Energy Accelerator Research Organization (KEK), Tsukuba, Ibaraki, Japan}}
\newcommand{\INSTED}{\affiliation{Institut de Fisica d'Altes Energies (IFAE), The Barcelona Institute of Science and Technology, Campus UAB, Bellaterra (Barcelona) Spain}}
\newcommand{\INSTEC}{\affiliation{IFIC (CSIC \& University of Valencia), Valencia, Spain}}
\newcommand{\INSTEI}{\affiliation{Imperial College London, Department of Physics, London, United Kingdom}}
\newcommand{\INSTGF}{\affiliation{INFN Sezione di Bari and Universit\`a e Politecnico di Bari, Dipartimento Interuniversitario di Fisica, Bari, Italy}}
\newcommand{\INSTBE}{\affiliation{INFN Sezione di Napoli and Universit\`a di Napoli, Dipartimento di Fisica, Napoli, Italy}}
\newcommand{\INSTBF}{\affiliation{INFN Sezione di Padova and Universit\`a di Padova, Dipartimento di Fisica, Padova, Italy}}
\newcommand{\INSTBD}{\affiliation{INFN Sezione di Roma and Universit\`a di Roma ``La Sapienza'', Roma, Italy}}
\newcommand{\INSTEB}{\affiliation{Institute for Nuclear Research of the Russian Academy of Sciences, Moscow, Russia}}
\newcommand{\INSTHA}{\affiliation{Kavli Institute for the Physics and Mathematics of the Universe (WPI), The University of Tokyo Institutes for Advanced Study, University of Tokyo, Kashiwa, Chiba, Japan}}
\newcommand{\INSTCC}{\affiliation{Kobe University, Kobe, Japan}}
\newcommand{\INSTCD}{\affiliation{Kyoto University, Department of Physics, Kyoto, Japan}}
\newcommand{\INSTEJ}{\affiliation{Lancaster University, Physics Department, Lancaster, United Kingdom}}
\newcommand{\INSTFC}{\affiliation{University of Liverpool, Department of Physics, Liverpool, United Kingdom}}
\newcommand{\INSTFI}{\affiliation{Louisiana State University, Department of Physics and Astronomy, Baton Rouge, Louisiana, U.S.A.}}
\newcommand{\INSTJ}{\affiliation{Universit\'e de Lyon, Universit\'e Claude Bernard Lyon 1, IPN Lyon (IN2P3), Villeurbanne, France}}
\newcommand{\INSTHB}{\affiliation{Michigan State University, Department of Physics and Astronomy,  East Lansing, Michigan, U.S.A.}}
\newcommand{\INSTCE}{\affiliation{Miyagi University of Education, Department of Physics, Sendai, Japan}}
\newcommand{\INSTDF}{\affiliation{National Centre for Nuclear Research, Warsaw, Poland}}
\newcommand{\INSTFJ}{\affiliation{State University of New York at Stony Brook, Department of Physics and Astronomy, Stony Brook, New York, U.S.A.}}
\newcommand{\INSTGJ}{\affiliation{Okayama University, Department of Physics, Okayama, Japan}}
\newcommand{\INSTCF}{\affiliation{Osaka City University, Department of Physics, Osaka, Japan}}
\newcommand{\INSTGG}{\affiliation{Oxford University, Department of Physics, Oxford, United Kingdom}}
\newcommand{\INSTBB}{\affiliation{UPMC, Universit\'e Paris Diderot, CNRS/IN2P3, Laboratoire de Physique Nucl\'eaire et de Hautes Energies (LPNHE), Paris, France}}
\newcommand{\INSTGC}{\affiliation{University of Pittsburgh, Department of Physics and Astronomy, Pittsburgh, Pennsylvania, U.S.A.}}
\newcommand{\INSTFA}{\affiliation{Queen Mary University of London, School of Physics and Astronomy, London, United Kingdom}}
\newcommand{\INSTE}{\affiliation{University of Regina, Department of Physics, Regina, Saskatchewan, Canada}}
\newcommand{\INSTGD}{\affiliation{University of Rochester, Department of Physics and Astronomy, Rochester, New York, U.S.A.}}
\newcommand{\INSTHC}{\affiliation{Royal Holloway University of London, Department of Physics, Egham, Surrey, United Kingdom}}
\newcommand{\INSTBC}{\affiliation{RWTH Aachen University, III. Physikalisches Institut, Aachen, Germany}}
\newcommand{\INSTFB}{\affiliation{University of Sheffield, Department of Physics and Astronomy, Sheffield, United Kingdom}}
\newcommand{\INSTDI}{\affiliation{University of Silesia, Institute of Physics, Katowice, Poland}}
\newcommand{\INSTEH}{\affiliation{STFC, Rutherford Appleton Laboratory, Harwell Oxford,  and  Daresbury Laboratory, Warrington, United Kingdom}}
\newcommand{\INSTCH}{\affiliation{University of Tokyo, Department of Physics, Tokyo, Japan}}
\newcommand{\INSTBJ}{\affiliation{University of Tokyo, Institute for Cosmic Ray Research, Kamioka Observatory, Kamioka, Japan}}
\newcommand{\INSTCG}{\affiliation{University of Tokyo, Institute for Cosmic Ray Research, Research Center for Cosmic Neutrinos, Kashiwa, Japan}}
\newcommand{\INSTGI}{\affiliation{Tokyo Metropolitan University, Department of Physics, Tokyo, Japan}}
\newcommand{\INSTF}{\affiliation{University of Toronto, Department of Physics, Toronto, Ontario, Canada}}
\newcommand{\INSTB}{\affiliation{TRIUMF, Vancouver, British Columbia, Canada}}
\newcommand{\INSTG}{\affiliation{University of Victoria, Department of Physics and Astronomy, Victoria, British Columbia, Canada}}
\newcommand{\INSTDJ}{\affiliation{University of Warsaw, Faculty of Physics, Warsaw, Poland}}
\newcommand{\INSTDH}{\affiliation{Warsaw University of Technology, Institute of Radioelectronics, Warsaw, Poland}}
\newcommand{\INSTFD}{\affiliation{University of Warwick, Department of Physics, Coventry, United Kingdom}}
\newcommand{\INSTGE}{\affiliation{University of Washington, Department of Physics, Seattle, Washington, U.S.A.}}
\newcommand{\INSTGH}{\affiliation{University of Winnipeg, Department of Physics, Winnipeg, Manitoba, Canada}}
\newcommand{\INSTEA}{\affiliation{Wroclaw University, Faculty of Physics and Astronomy, Wroclaw, Poland}}
\newcommand{\INSTHE}{\affiliation{Yokohama National University, Faculty of Engineering, Yokohama, Japan}}
\newcommand{\INSTH}{\affiliation{York University, Department of Physics and Astronomy, Toronto, Ontario, Canada}}

\INSTEE
\INSTFE
\INSTD
\INSTGA
\INSTI
\INSTGB
\INSTFG
\INSTFH
\INSTBA
\INSTEF
\INSTEG
\INSTDG
\INSTCB
\INSTED
\INSTEC
\INSTEI
\INSTGF
\INSTBE
\INSTBF
\INSTBD
\INSTEB
\INSTHA
\INSTCC
\INSTCD
\INSTEJ
\INSTFC
\INSTFI
\INSTJ
\INSTHB
\INSTCE
\INSTDF
\INSTFJ
\INSTGJ
\INSTCF
\INSTGG
\INSTBB
\INSTGC
\INSTFA
\INSTE
\INSTGD
\INSTHC
\INSTBC
\INSTFB
\INSTDI
\INSTEH
\INSTCH
\INSTBJ
\INSTCG
\INSTGI
\INSTF
\INSTB
\INSTG
\INSTDJ
\INSTDH
\INSTFD
\INSTGE
\INSTGH
\INSTEA
\INSTHE
\INSTH

\author{K.\,Abe}\INSTBJ
\author{J.\,Amey}\INSTEI
\author{C.\,Andreopoulos}\INSTEH\INSTFC
\author{M.\,Antonova}\INSTEB
\author{S.\,Aoki}\INSTCC
\author{A.\,Ariga}\INSTEE
\author{D.\,Autiero}\INSTJ
\author{S.\,Ban}\INSTCD
\author{M.\,Barbi}\INSTE
\author{G.J.\,Barker}\INSTFD
% ...

\collaboration{The T2K Collaboration}\noaffiliation


\date{\today}

\begin{abstract}

T2K reports its first results in the search for CP violation in neutrino
oscillations using appearance and disappearance channels for neutrino- and
antineutrino-mode beam.  The data include all runs from Jan 2010 to May 2016
and comprise $7.482\times10^{20}$\,protons on target in neutrino mode, which yielded in the far detector
32 e-like and 135 $\mu$-like events, and
$7.471\times10^{20}$\,protons on target in antineutrino mode which yielded 4 e-like and
66 $\mu$-like events. Reactor measurements of
$\sin^{2}2\theta_{13}$ have been used as an additional constraint.
The one-dimensional confidence interval at 90\% for $\delta_{CP}$ spans the
range ($-3.13$, $-0.39$) for normal mass ordering.
The CP conservation hypothesis ($\delta_{CP}=0,\pi$)
is excluded at 90\% C.L.

\end{abstract}


\maketitle

\noindent \textit{Introduction --- }
A new source of CP violation beyond the CKM quark mixing matrix is necessary to explain observations of baryon asymmetry in the Universe.
In the lepton sector the PMNS framework~\cite{Maki:1962mu,Pontecorvo:1967fh} allows for CP violation. The first indication of non-zero $\theta_{13}$\cite{abe2011} followed by its discovery \cite{PhysRevLett.108.171803, PhysRevLett.108.191802, PhysRevLett.108.131801} and then the discovery of \numunue oscillation by T2K~\cite{Abe:2013hdq} have opened the possibility to look for 
CP violation in neutrino oscillation.

\begin {figure}[htbp]
  \begin{center}
    \includegraphics[width=0.45\textwidth]{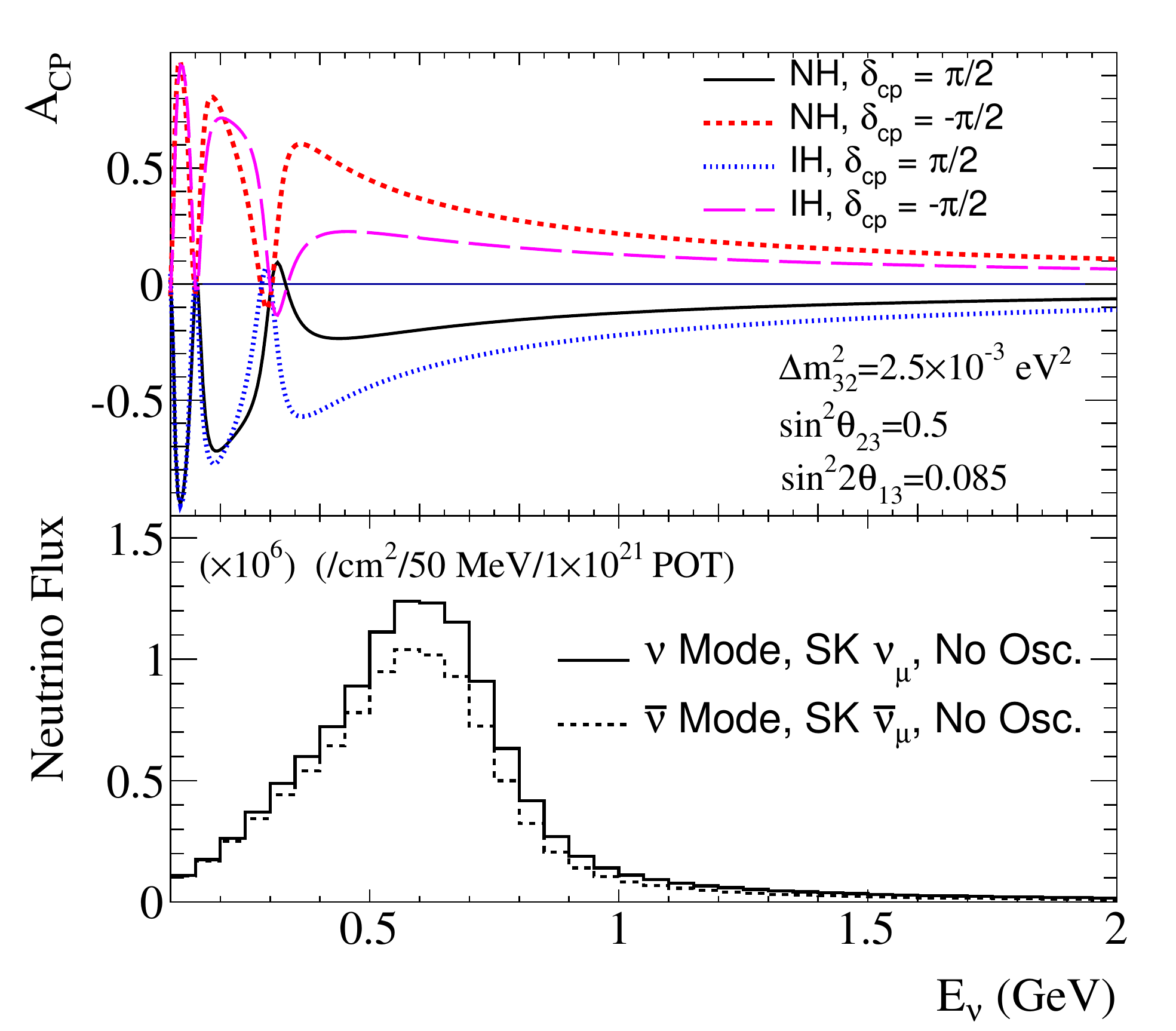}
    \caption{The leptonic CP asymmetry, $A_{CP} = [P(\nu_{\mu} \rightarrow \nu_e)-P(\bar{\nu}_{\mu} \rightarrow \bar{\nu}_e)]/[P(\nu_{\mu} \rightarrow \nu_e)+P(\bar{\nu}_{\mu} \rightarrow \bar{\nu}_e)]$, as a function of energy for maximal CP-violation hypotheses (top) and the \numu (\numubar) component of the unoscillated (anti)neutrino flux in neutrino and antineutrino modes (bottom).}
    \label{fig:flux_cpasymm}
  \end{center}
\end {figure}

In this Letter we present the first joint fit of neutrino and antineutrino $\bbar{\nu}_{\mu} \rightarrow \bbar{\nu}_e$ and $\bbar{\nu}_{\mu} \rightarrow \bbar{\nu}_{\mu}$ oscillation at T2K.
The mixing of neutrinos in the three-flavour framework is represented by the unitary PMNS matrix, parameterized by three mixing angles, $\theta_{12}$, $\theta_{13}$, and $\theta_{23}$, and a CP-violating phase $\delta_{CP}$~\cite{PDG2015}. The 
 probability for $\bbar{\nu}_{\mu} \rightarrow \bbar{\nu}_e$ oscillation,
 as a function of neutrino propagation distance $L$ and energy $E$, can be written:
\begin{alignat}{3}
P(\bbar{\nu}_{\mu} \rightarrow \bbar{\nu}_e) &\simeq 
	&&\sin^2\theta_{23} \sin^22\theta_{13} \sin^2\frac{\Delta m^2_{31} L}{4E} \nonumber \\ 
	&\bplus{-}&&\frac{\sin2\theta_{12} \sin2\theta_{23}}{2\sin\theta_{13}} \sin\frac{\Delta m^2_{21} L}{4E} \nonumber \\
	&&&\times \sin^22\theta_{13} \sin^2\frac{\Delta m^2_{31} L}{4E} \sin\delta_{CP} \nonumber \\
	&+&&\mbox{(CP-even, solar, matter effect terms)}
\label{eq:numu-nue}
\end{alignat}

\noindent where $\Delta m^2_{ij}=m^2_{i}-m^2_{j}$ represents the neutrino mass-squared difference between mass eigenstates $i$ and $j$. 
The $\bbar{\nu}_{\mu}~\rightarrow~\bbar{\nu}_{\mu}$ survival probability is dominated by the parameters $\sin^2\theta_{23}$ and $\Delta m^2_{32}$, as given in~\cite{PhysRevLett.111.211803}.
Comparing electron neutrino and antineutrino appearance probabilities allows a direct measurement of 
CP violation at T2K.  The asymmetry variable ($A_{CP}=P({\nu}_{\mu} \rightarrow{\nu}_e)-P(\bar{\nu}_{\mu} \rightarrow \bar{\nu}_e))/(P({\nu}_{\mu} \rightarrow{\nu}_e)+P(\bar{\nu}_{\mu} \rightarrow \bar{\nu}_e))$ and the \numu (\numubar) component of the expected T2K flux without oscillations are shown in Fig.~\ref{fig:flux_cpasymm}. At the flux peak energy, $A_{CP}$ can be as large as 0.4
, including a contribution of around 0.1 due to matter effects.

\noindent \textit{The T2K Experiment --- }
The T2K experiment~\cite{Abe:2011ks} uses a 30~GeV proton beam from the J-PARC accelerator facility to produce a muon (anti)neutrino beam. The proton beam strikes a graphite target to produce charged pions and kaons, which are focused by three magnetic horns. Depending on the polarity of the horn current, either positively- or negatively-charged mesons are focused, resulting in a beam largely composed of muon neutrinos or antineutrinos. A 96-m decay volume lies downstream of the magnetic horns, followed by the beam dump and muon monitor~\cite{Suzuki:2014jyd}. The neutrino beam is measured by detectors placed on axis and off axis at 2.5$^{\circ}$ relative to the beam direction. The off-axis neutrino energy spectrum peaks at 0.6~GeV, and has a reduced $\bbar{\nu}_e$ contamination and smaller backgrounds from higher energy neutrinos than the on-axis spectrum. Two detectors located 280~m from the target are used to measure the beam direction, spectrum, and composition, as well as the event rate: INGRID (on axis)~\cite{Abe:2011xv}, and ND280 (off axis), which is housed inside a 0.2~T magnet. The Super-Kamiokande (Super-K) 50-kt water Cherenkov detector~\cite{fukuda}, located off axis and 295~km from the neutrino production point, is used to detect oscillated neutrinos.
	
\noindent \textit{Data Sets --- }
The results presented here are based on data collected in two periods: one in which the beam operated solely in neutrino mode, January 2010 --- May 2013, and one in which the beam operated mostly in antineutrino mode, May 2014 --- May 2016.
This corresponds to a neutrino beam exposure of $7.482 \times 10^{20}$ protons on target (POT) in neutrino mode and $7.471 \times 10^{20}$ POT in antineutrino mode for the far detector analysis, and an exposure of $5.82 \times 10^{20}$ POT in neutrino mode and $2.84 \times 10^{20}$ POT in antineutrino mode for the near detector analysis.

\noindent \textit{Analysis Strategy --- }
The analysis strategy is similar to that of previous T2K results~\cite{PhysRevLett.111.211803,Abe:2013hdq,Abe:2015awa,Abe:2015ibe}: oscillation parameters are estimated by comparing predictions and observations at the far detector. A tuned prediction of the oscillated spectrum at the far detector, with associated uncertainty, is obtained by fitting samples of charged-current interactions at ND280. The analysis presented here differs from previous results in that both neutrino and antineutrino samples are fitted at both ND280 and Super-K. Including antineutrino data at ND280 ensures that the interaction model is consistent between neutrinos and antineutrinos and provides a constraint on the wrong-sign background in the antineutrino-mode beam.

\noindent \textit{Neutrino Flux Model --- }
The T2K neutrino and antineutrino fluxes at near and far detectors, and their correlations, are calculated~\cite{Abe:2012av} using a data-driven
hybrid simulation with FLUKA 2011~\cite{cite:FLUKA1,Ferrari:2005zk} used to simulate hadronic
interactions and transport particles inside the target, while GEANT3~\cite{Brun:1994aa} with
GCALOR~\cite{Zeitnitz:1994bs} is used to simulate the rest of the neutrino beam line.  
The interactions
of hadrons in both FLUKA 2011 and GCALOR are tuned using thin target hadron production data,
including measurements of the total cross section for particle production, and $\pi^{\pm}$, $K^{\pm}$, $p^{+}$, $\Lambda$ and $K^{0}_{S}$ production
with 30~GeV protons on a graphite target by the NA61/SHINE experiment~\cite{Abgrall:2015xoa}.
The uncertainty on the flux calculation is estimated by propagating systematic variations through
the flux calculation procedure.  Dominant systematic error sources include uncertainties on
the NA61/SHINE hadron production measurements, hadronic interaction length measurements from NA61/SHINE and other experiments,
the initial proton beam trajectory and the horn currents.  The total uncertainty on the flux
near the peak energy is $\sim9\%$.  The \numu (\numubar) component of the predicted fluxes without oscillations are shown in Fig.~\ref{fig:flux_cpasymm}. 
At the far detector and in the absence of oscillations, we predict that 94.1\% (92.3\%) of the T2K neutrino-mode (antineutrino-mode) beam below 1.25~GeV is \numu (\numubar).
The $\bar{\nu}_{\mu}$ flux in antineutrino mode is reduced by $\sim20\%$ relative
to the $\nu_{\mu}$ flux in neutrino mode due to the smaller production cross section for $\pi^{-}$
relative to $\pi^{+}$ in 30~GeV $p+C$ interactions.


\noindent \textit{Neutrino Interaction Model --- }
The interactions of neutrinos and antineutrinos with nuclei in the near and far detectors
are modelled with the NEUT~\cite{Hayato:2009zz} neutrino interaction generator.
The charged-current quasielastic (CCQE) interactions are modelled with a 
relativistic Fermi gas (RFG) nuclear model with relativistic corrections for
long range correlations using the random phase approximation (RPA) as applied
by Nieves {\it et al.}~\cite{Nieves:2004wx}. The choice of CCQE nuclear model
was made based on fits to external CCQE-like data~\cite{NIWGpaper} from the
MiniBooNE~\cite{AguilarArevalo:2010zc,AguilarArevalo:2013hm} 
and MINERvA~\cite{Fiorentini:2013ezn,Fields:2013zhk} experiments.
Interactions on more than one nucleon are modelled with an implementation of the
2p-2h model developed by Nieves {\it et al.}~\cite{Nieves:2011pp,Gran:2013kda}.
These interactions are characterized by multi-nucleon ejection and no final state pions;
hence they may be confused for CCQE interactions in a water Cherenkov detector.
The single pion production model in NEUT has been tuned using form factors
from Graczyk and Sobczyk~\cite{Graczyk:2007bc} and with a reanalysis of ANL and BNL 
bubble chamber data sets~\cite{Wilkinson:2014yfa}.
The coherent pion production model has been tuned to reproduce data from MINERvA \cite{minervacoh} and T2K \cite{t2kcoh}.
At the T2K peak energy, the antineutrino cross section is $\sim3.5$ times smaller than 
the neutrino cross section. 

The parameterization of uncertainties in the neutrino interaction model is largely
unchanged from previous measurements~\cite{Abe:2015ibe,Abe:2015awa}.  
Parameters that vary the binding energy, Fermi momentum, 2p-2h normalization and
charged current coherent cross-section normalization are applied separately for interactions on
carbon and oxygen.
To cover the different predictions by Nieves {\it et al.}~\cite{Nieves:2011pp,Gran:2013kda} and 
Martini {\it et al.}~\cite{Martini:2009uj,Martini:2013sha} of the relative 2p-2h interaction rates for
neutrinos and antineutrinos, the normalizations of 2p-2h interactions for neutrinos and antineutrinos
are allowed to vary independently.  

Only the interactions of $\nu_{\mu}$ and $\bar{\nu}_{\mu}$ are explicitly constrained by near detector measurements
in this analysis.
Since the oscillation signals include $\nu_{e}$ and $\bar{\nu}_{e}$ interactions,
it is necessary to assign uncertainties on the cross section ratios $\sigma_{\nu_{e}}/\sigma_{\nu_{\mu}}$
and $\sigma_{\bar{\nu}_{e}}/\sigma_{\bar{\nu}_{\mu}}$.  
Following the treatment in~\cite{Day:2012gb}, separate parameters
for $\sigma_{\nu_{e}}/\sigma_{\nu_{\mu}}$ and $\sigma_{\bar{\nu}_{e}}/\sigma_{\bar{\nu}_{\mu}}$ are introduced
with a theoretical uncertainty of 2.8\% for each.  A correlation coefficient of -0.5 is assumed for these two parameters,
accounting for anti-correlated changes to the relative cross section rates that can arise from nucleon form-factor
variations.

\noindent \textit{Fit to Near Detector Data --- }
The systematic parameters in the neutrino flux and interaction models
are constrained by a fit to charged current (CC) candidate samples in 
the ND280~\cite{Abe:2011ks} near detector.  The data sets used consist of reconstructed
interactions in two fine-grained detectors (FGDs)~\cite{Amaudruz:2012pe} with particle
tracking in three time projection chambers (TPCs)~\cite{Abgrall:2010hi}.  
FGD2 contains six 2.54-cm-thick water panels,
allowing systematic parameters governing neutrino interactions on H$_{2}$O, the 
same target as Super-K, to be directly constrained.
The CC candidate samples in ND280 are divided into 
categories based on the beam mode (neutrino vs. antineutrino), the FGD
in which the interaction takes place, the muon charge and the final state 
multiplicity.  For data taken in neutrino mode, only interactions with a
negatively charged muon are considered.
For data taken in antineutrino mode, there are separate categories for 
events with positively-charged (right-sign) and negatively-charged (wrong-sign)
muon candidates.  The wrong-sign candidates are included because the larger
neutrino cross section leads to a non-negligible wrong-sign background in 
antineutrino mode.  In neutrino mode, there are three categories for 
reconstructed final states: no pion candidate in the final state (CC0$\pi$), one 
pion candidate in the final state (CC1$\pi$) and all other CC candidates (CC Other).  In
antineutrino mode, events are divided into two categories based on the final
states: only the muon track exits the FGD to enter the TPC (CC 1-Track) and at least 
one other track enters the TPC (CC N-Track).  

When fitting, the data
are binned according to the momentum of the muon candidate, $p_{\mu}$ and $\textrm{cos}\theta_{\mu}$,
where $\theta_{\mu}$ is the angle of the muon direction 
relative to
the central axis of the detector, roughly 1.7$^{\circ}$ away from the
incident (anti)neutrino direction.
A binned maximum likelihood fit is performed
in which the neutrino flux and interaction model parameters are allowed to vary.
Nuisance parameters describing the systematic errors in the ND280 detector model -- the largest of which is pion interaction modelling -- are marginalised in the fit.

\begin {figure}[htbp]
  \begin{center}
    \includegraphics[width=0.48\textwidth]{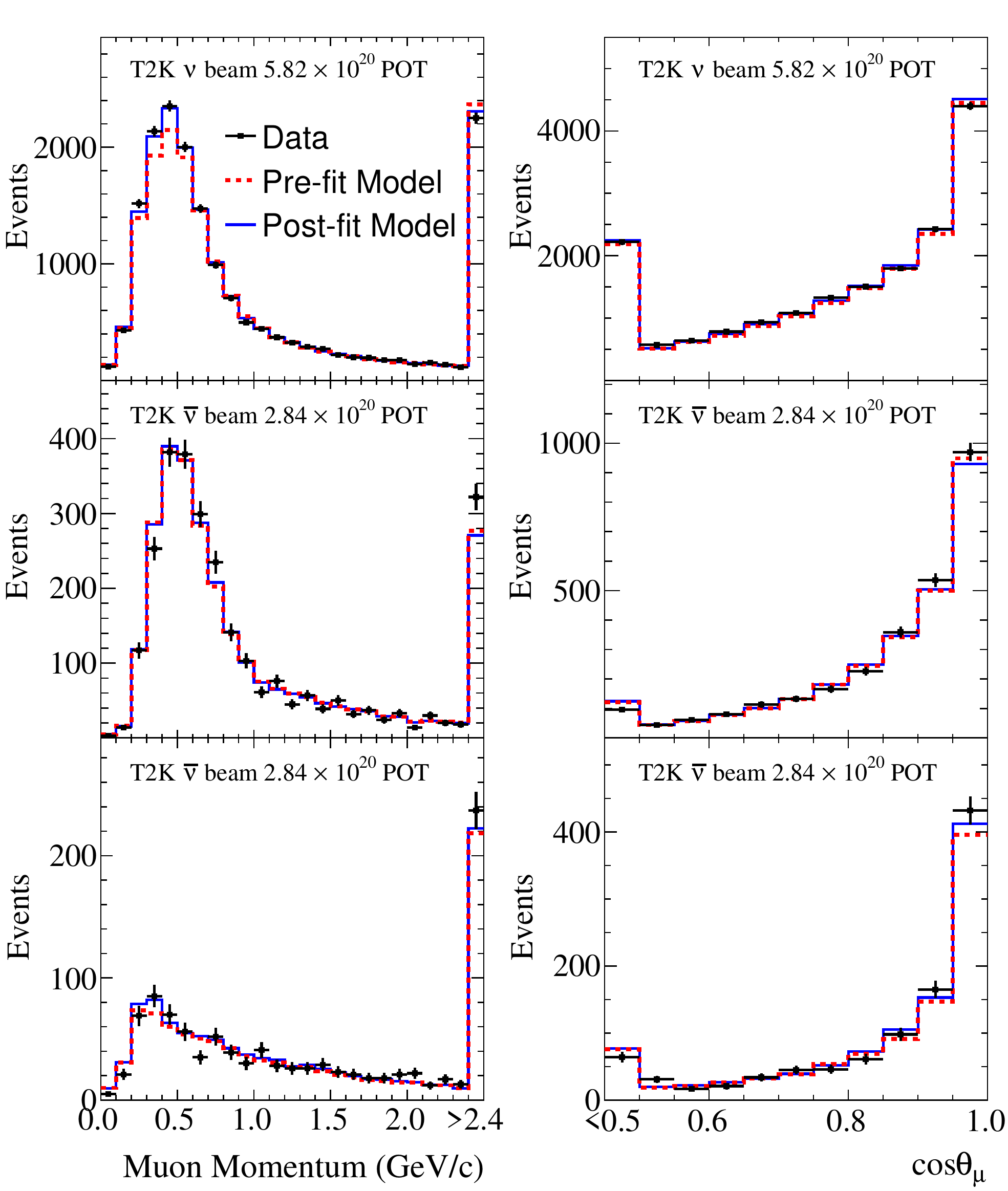}
    \caption{The FGD2 data, pre-fit predictions and post-fit predictions binned in $p_{\mu}$ (left) and $\textrm{cos}\theta_{\mu}$ (right) for
the neutrino mode CC0$\pi$ (top), antineutrino mode CC 1-Track $\mu^{+}$ (middle) and  antineutrino mode CC 1-Track $\mu^{-}$ (bottom) categories.
The overflow bins are integrated out to 10000~MeV/c and -1.0 for $p_{\mu}$  and the $\textrm{cos}\theta_{\mu}$ respectively.}
    \label{fig:nd280_postfit}
  \end{center}
\end {figure}

The fitted $p_{\mu}$ and $\textrm{cos}\theta_{\mu}$ distributions for the FGD2 CC0$\pi$ and CC~1-Track
categories are shown in Fig.~\ref{fig:nd280_postfit}.  
Acceptable agreement between the post-fit model and data is observed for
both kinematic variables,  
with a p-value of 0.086.
The best-fit fluxes are increased with respect to the original flux model by 10-15\% near the flux peak. This is driven
by the pre-fit deficit in the prediction for the CC0$\pi$ and CC Other samples. The fitted value for the axial mass in the CCQE model is 1.12~GeV/$c^{2}$, compared to 1.24~GeV/$c^{2}$ in a previous fit
where the 2p-2h model and RPA corrections were not included~\cite{Abe:2015awa}.  The fit to ND280 data 
reduces the 
uncertainty on the event-rate predictions at the far detector due to uncertainties on the flux and ND280-constrained interaction model parameters from 10.9\%(12.4\%) to 2.9\%(3.2\%) for the $\nu_{e}$($\bar{\nu}_{e}$) candidate sample.

\begin{figure}[htb]
\begin{minipage}[c]{0.49\linewidth}
\includegraphics[width=\textwidth]{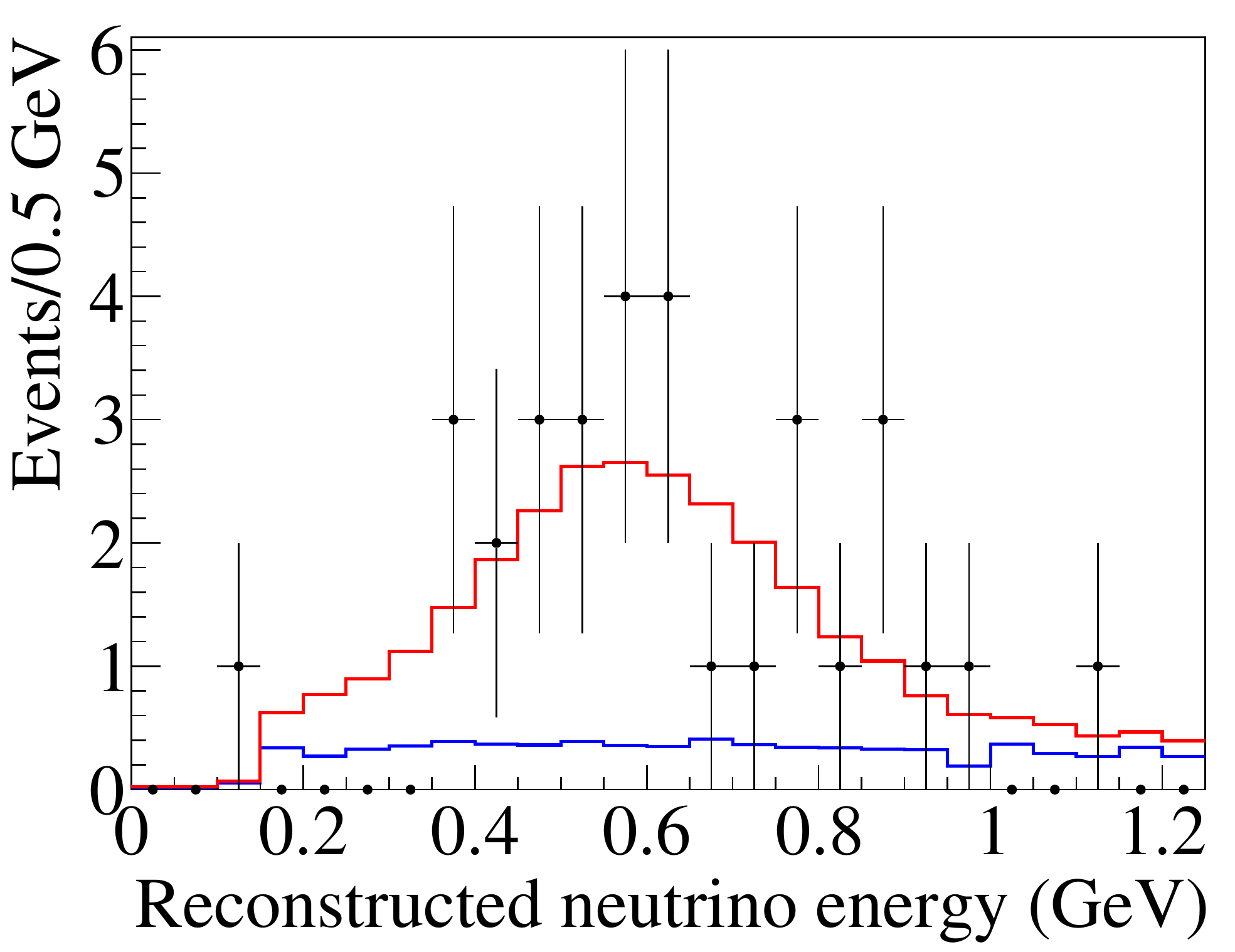}
\end{minipage} \hfill
\begin{minipage}[c]{0.49\linewidth}
\includegraphics[width=\textwidth]{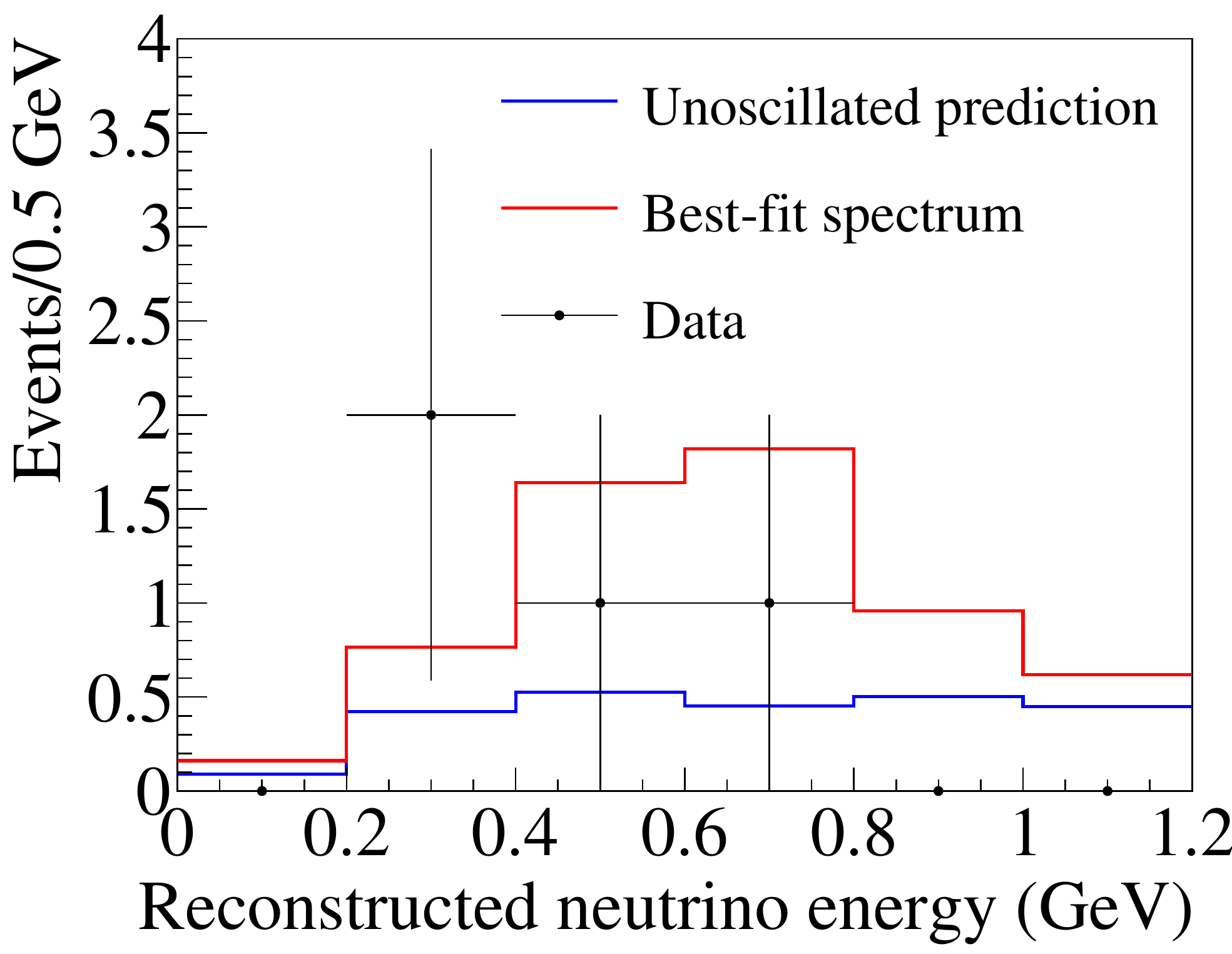}
\end{minipage}
\caption{
The reconstructed neutrino energy at the far detector for the $\nu_e$ (left) and $\bar{\nu}_e$ (right) candidate samples 
is shown together with the expected distribution without oscillation (blue histogram) and the best fit (red histogram).
}
\label{fig:spectra_bf}
\end{figure}

\noindent \textit{Far Detector Data --- }
At the far detector, events are extracted that lie within $[-2,10]\,\mu$s
relative to the beam arrival.
Fully contained events within the fiducial volume are selected by requiring that no
hit cluster is observed in the outer detector volume,
that the distance from the reconstructed vertex to the inner detector wall is larger than 2\,m, and that the total
observed charge is greater than the
equivalent quantity for a 30\,MeV electron.
The CCQE component of our sample is enhanced by selecting events with
a single Cherenkov ring.
The \numu/\numubar CCQE candidate samples are then selected by requiring
a $\mu$-like ring using a PID-likelihood, zero or one
decay electron candidates and muon momentum greater than 200\,MeV/c
to reduce pion background.
Post selection, 135 and 66 events remain in the \numu and \numubar
candidate samples respectively, while
if 
$|\Delta m^2_{32}| = 2.509 \times 10^{-3}~\rm{eV}^2/\rm{c}^4 $
and
$\sin^2 \theta_{23} = 0.528$ (i.e. maximal disappearance),
 135.5 and 64.1 events are expected.
The \nue/\nuebar CCQE candidate samples are selected by requiring an e-like
ring, zero decay electron candidates, not $\pi^0$-like and
reconstructed energy less than 1.25\,GeV.
The total number of events remaining in these samples is presented in Table \ref{tab:evtnue}
with their respective expectation for different values of $\delta_{CP}$,
$\sin^2 2\theta_{13} = 0.085$, $|\Delta m^2_{32}| = 2.509 \times 10^{-3}~\rm{eV}^2/\rm{c}^4 $,
and
$\sin^2 \theta_{23} = 0.528$. 
The $\nu_e$ ($\bar{\nu}_e$) contamination in the 
\nuebar (\nue) sample is 17.4 (0.5) \%, and the proportion of the sample expected to correspond to oscillated \nuebar (\nue) events is 46.4 (80.9) \% for $\delta_{CP}=-\pi/2$.
A more detailed description of the candidate event selections can be found in previous publications~\cite{Abe:2015awa}.
The reconstructed neutrino energy spectra for the $\nu_e$ and $\bar{\nu}_e$ samples is shown in 
Fig.~\ref{fig:spectra_bf}.
The \nuebar signal events are
concentrated in the forward direction with respect to the beam, unlike the backgrounds.
Therefore, incorporating reconstructed lepton angle information in the analysis increases the sensitivity.

The systematic errors concerning the detector behaviour are
estimated using atmospheric neutrino and cosmic-ray muon
events.  A sample of hybrid data-MC events is also used to
evaluate uncertainties regarding $\pi^0$ rejection.
Correlations between the uncertainties for the four samples are
considered.

\begin{table}[htbp]
    \centering
    \caption{Number of $\nu_e$ and $\overline{\nu}_e$ events expected for various values of $\delta_{CP}$ and both mass orderings 
    compared to the observed numbers.}
    \label{tab:evtnue}
    \begin{tabular}{|c|c|c|c|c|c|}
        \hline
        Normal & $\delta_{CP}= -\pi/2$ & $\delta_{CP}= 0 $ & $\delta_{CP}= \pi/2$ &  $\delta_{CP}= \pi$  & Observed\\
        \hline 
        $\nu_e$ &   28.7 & 24.2& 19.6& 24.1& 32 \\
        $\overline{\nu}_e$ &  6.0 &6.9& 7.7 &6.8 &4 \\     
        \hline
        \hline
        Inverted & $\delta_{CP}= -\pi/2$ & $\delta_{CP}= 0 $ & $\delta_{CP}= \pi/2$ &  $\delta_{CP}= \pi$  & Observed\\
        \hline 
        $\nu_e$ 			& 25.4 	& 21.3	& 17.1	& 21.3	& 32 \\
        $\overline{\nu}_e$ 	& 6.5 	& 7.4		& 8.4		& 7.4		&4 \\    
\hline
    \end{tabular}
\end{table}

The fractional variation of the number of expected events for the four samples
owing to the various sources of systematic uncertainty 
are shown in Table~\ref{tab:fracevent}. A more in-depth description of the sources of systematic uncertainty in the fit is given in~\cite{Abe:2015awa}, although this reference does not cover the updates discussed in previous sections.

\begin{table}[htbp]
    \centering
    \caption{
Systematic uncertainty on the predicted event rate at the far detector.
    }
    \label{tab:fracevent}
    \begin{tabular}{|c|c|c|c|c|}
        \hline
        Source [\%] 							& $\nu_\mu$ 	& $\nu_e$ 	& $\overline{ \nu}_\mu$ 	& $\overline{\nu}_e$ \\
        \hline    
        ND280-unconstrained cross section   		& 0.7 		& 3.0 		& 0.8 				& 3.3  \\
        Flux and ND280-constrained cross section    	& 2.8			& 2.9 		& 3.3 				& 3.2 \\
        SK detector systematics					& 3.9 		& 2.4 		& 3.3 				& 3.1 \\
        Final or secondary hadron interactions		& 1.5 		& 2.5 		& 2.1 				& 2.5 \\
        	\hline
Total & 5.0&  5.4&  5.2&  6.2 \\
\hline
\end{tabular}
\end{table}

\noindent \textit{Oscillation Analysis --- }
The oscillation parameters $\sin^2 \theta_{23}$,  
$\Delta m^2_{32}$, $\sin^2 \theta_{13}$ and  $\delta_{CP}$
are estimated
by performing a joint maximum-likelihood fit of the four far detector samples.
The oscillation probabilities are calculated using the full three-flavor oscillation formulae~\cite{Barger:1980tf}.  
Matter effects are included with an Earth density of $\rho = 2.6~g/\rm{cm}^3$~\cite{Hagiwara:2011kw}.

As described previously, the priors for the beam flux and neutrino interaction cross-section parameters 
are obtained from the fit with the near detector data.
The priors~\cite{PDG2015} for the solar neutrino oscillation parameters -- whose impact is almost negligible -- are 
$\sin^2 2\theta_{12} = 0.846 \pm 0.021 $, $\Delta m^2_{21} = (7.53 \pm 0.18) \times 10^{-5} ~\rm{eV}^2 / \rm{c}^4$,
and in some fits we use $\sin^2 2\theta_{13}=0.085 \pm 0.005$~\cite{PDG2015},
called the ``reactor measurement". 
Flat priors are used for $\sin^2 \theta_{23}$, $\Delta m^2_{32}$, and $\delta_{CP}$.

We use a procedure analogous to~\cite{Abe:2015ibe}:
after integrating over the prior distributions of the nuisance parameters a marginal likelihood, 
that depends only on the relevant oscillation parameters, is obtained.
We define
$-2 \Delta \ln \mathcal{L} = -2 \ln [ \mathcal{L}(\textbf{o}) / \mathcal{L}_{max} ]$
as the ratio between the marginal likelihood at the point $\textbf{o}$ of the relevant oscillation parameter space and the maximum marginal likelihood.

We have conducted three analyses using different far detector event 
quantities and different statistical approaches. 
All of them use the neutrino energy reconstructed in the CCQE hypothesis ($\rm{E}_{rec}$) for the $\bbar{\nu}_{\mu}$ samples.
The first analysis uses $\rm{E}_{rec}$ and the reconstructed angle between the lepton and the neutrino beam direction, $\theta_{lep}$,
 of the $\bbar{\nu}_{e}$ candidate samples
and provides confidence intervals using a hybrid Bayesian-frequentist approach~\cite{Cousins:1991qz}.
These results are shown in the following figures.
The second analysis is fully Bayesian and uses the lepton momentum, $p_{lep}$, and $\theta_{lep}$ for the $\bbar{\nu}_{e}$ samples
to compute credible intervals using the posterior probability.
The third analysis uses only $\rm{E}_{rec}$ spectra for the $\bbar{\nu}_{e}$ samples 
and a Markov Chain Monte Carlo method~\cite{Hastings:1970aa}
to provide Bayesian credible intervals.
This analysis performs a simultaneous fit of both the near and far detector data,
providing a validation of the extrapolation of the flux, cross section and detector systematic parameters from the near to far detector.
All three methods are in good agreement.

An indication of the sensitivity to $\delta_{CP}$ and the mass ordering can be obtained from Table~\ref{tab:evtnue}.
If CP violation is maximal ($\delta_{CP}=\pm \pi/2$), 
the predicted variation of the total number of events with respect to the CP conservation hypothesis ($\delta_{CP}=0, \pi$)
is about 20\%.
The different mass orderings induce a variation of the number of expected events of about 10\%.

\begin{figure}
\centering
\includegraphics[angle=0,width=9cm]{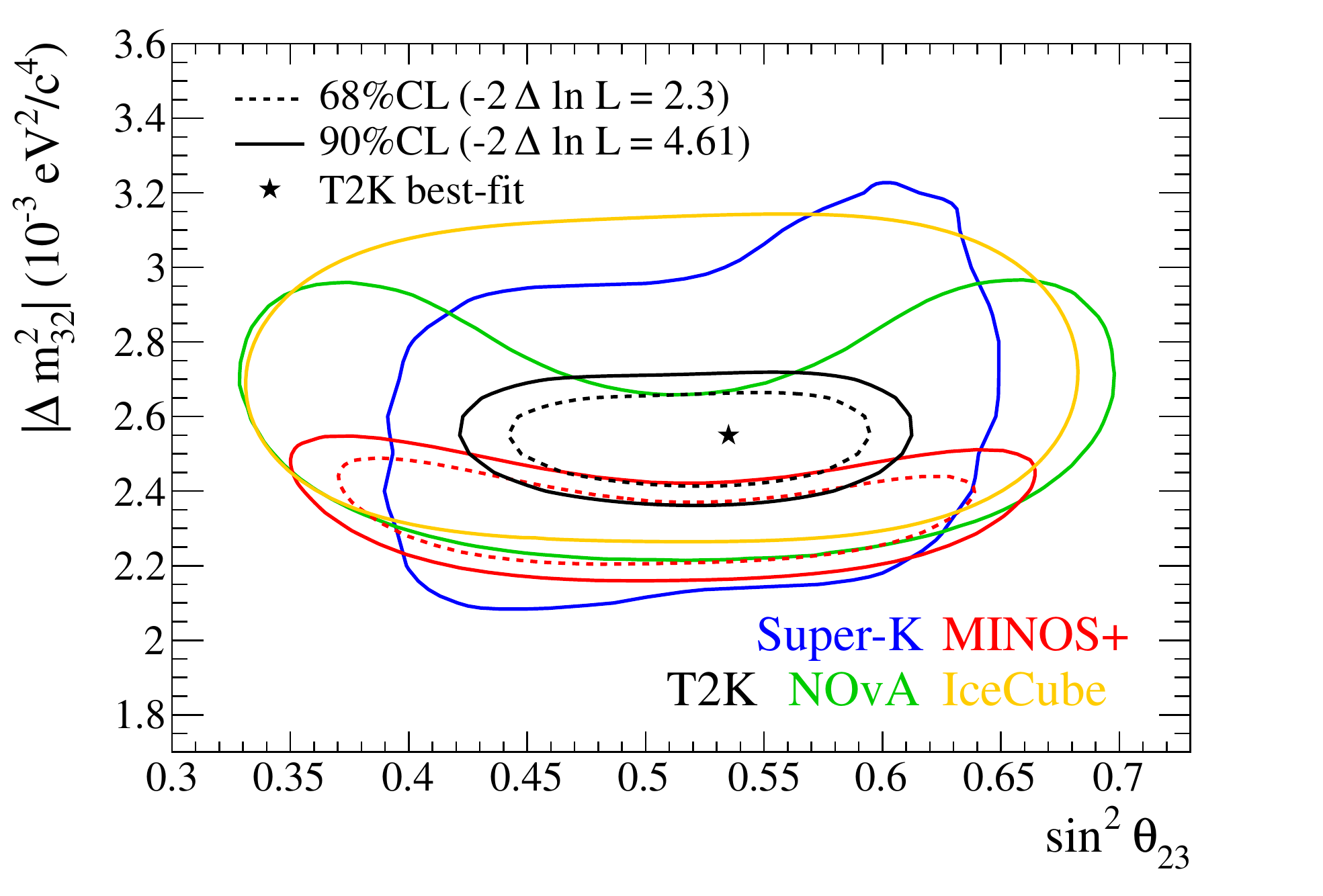}
\caption{
The 68\% (90\%) constant $-2 \Delta \ln L$ confidence regions for the 
$\sin^2 \theta_{23}$ $-$ $|\Delta m_{32}^2|$ plane assuming normal ordering, 
alongside NOvA\cite{nova2016}, MINOS+\cite{adamson}, SK\cite{wendell}, and IceCube\cite{aartsen} confidence regions.}
\label{fig:dm32theta23}
\end{figure}

A series of fits are performed where one or two oscillation parameters are determined and the others are marginalised.
Confidence regions are set using the constant $-2 \Delta \ln L$ method~\cite{PDG2015}. 
In the first fit 
confidence regions in the  $\sin^2 \theta_{23}$ $-$ $|\Delta m_{32}^2|$ plane (Fig.~\ref{fig:dm32theta23})
were computed 
using the reactor measurement of
$\sin^2 \theta_{13}$. 
The best-fit values are $\sin^2 \theta_{23}=0.532$ and $|\Delta m_{32}^2|= 2.545 \times 10^{-3}~\rm{eV}^2/\rm{c}^4$ 
($\sin^2 \theta_{23}=0.534$ and $|\Delta m_{32}^2|= 2.510 \times 10^{-3}~\rm{eV}^2/\rm{c}^4$)
for the normal (inverted) ordering.
The result is consistent with maximal disappearance. 
The T2K data weakly prefer the second octant ($\sin^2 \theta_{23}>0.5$)  
with a posterior probability of 61\%.

\begin{figure}
\centering
\includegraphics[angle=0,width=9cm]{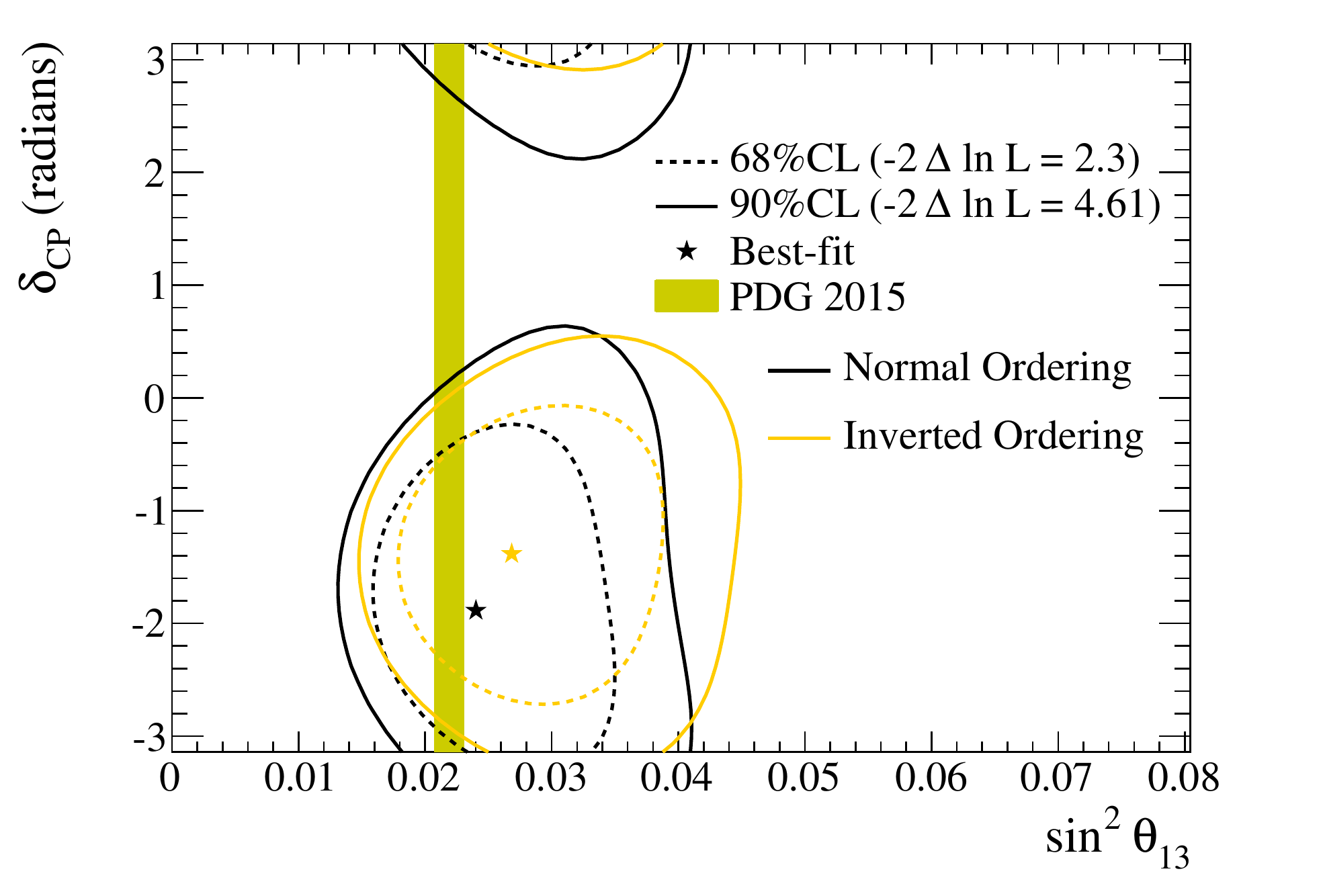}
\caption{
The 68\% (90\%) constant $-2 \Delta \ln L$ confidence regions in the $\delta_{CP}$ $-$ $\sin^2\theta_{13}$ plane are shown 
by the dashed (continuous) lines, computed independently 
for the normal (black) and inverted (red) mass ordering. 
The best-fit point is shown by a star for each mass ordering hypothesis. 
The 68\% confidence region from reactor experiments on $\sin^2 \theta_{13}$ is shown by the yellow vertical band.
}
\label{fig:th13delta}
\end{figure}

Confidence regions in the $\sin^2 \theta_{13}$ $-$ $\delta_{CP}$ plane are computed 
independently 
for both mass ordering hypotheses (Fig.~\ref{fig:th13delta})
without using the reactor measurement. 
The addition of antineutrino samples at Super-K gives the first sensitivity to $\delta_{CP}$ from T2K data alone.
There is good agreement between the T2K result and the reactor measurement for $\sin^2 \theta_{13}$. 
For both mass-ordering hypotheses, the best-fit value of $\delta_{CP}$ is close to $-\pi/2$.

\begin{figure}
\centering
\includegraphics[angle=0,width=9cm]{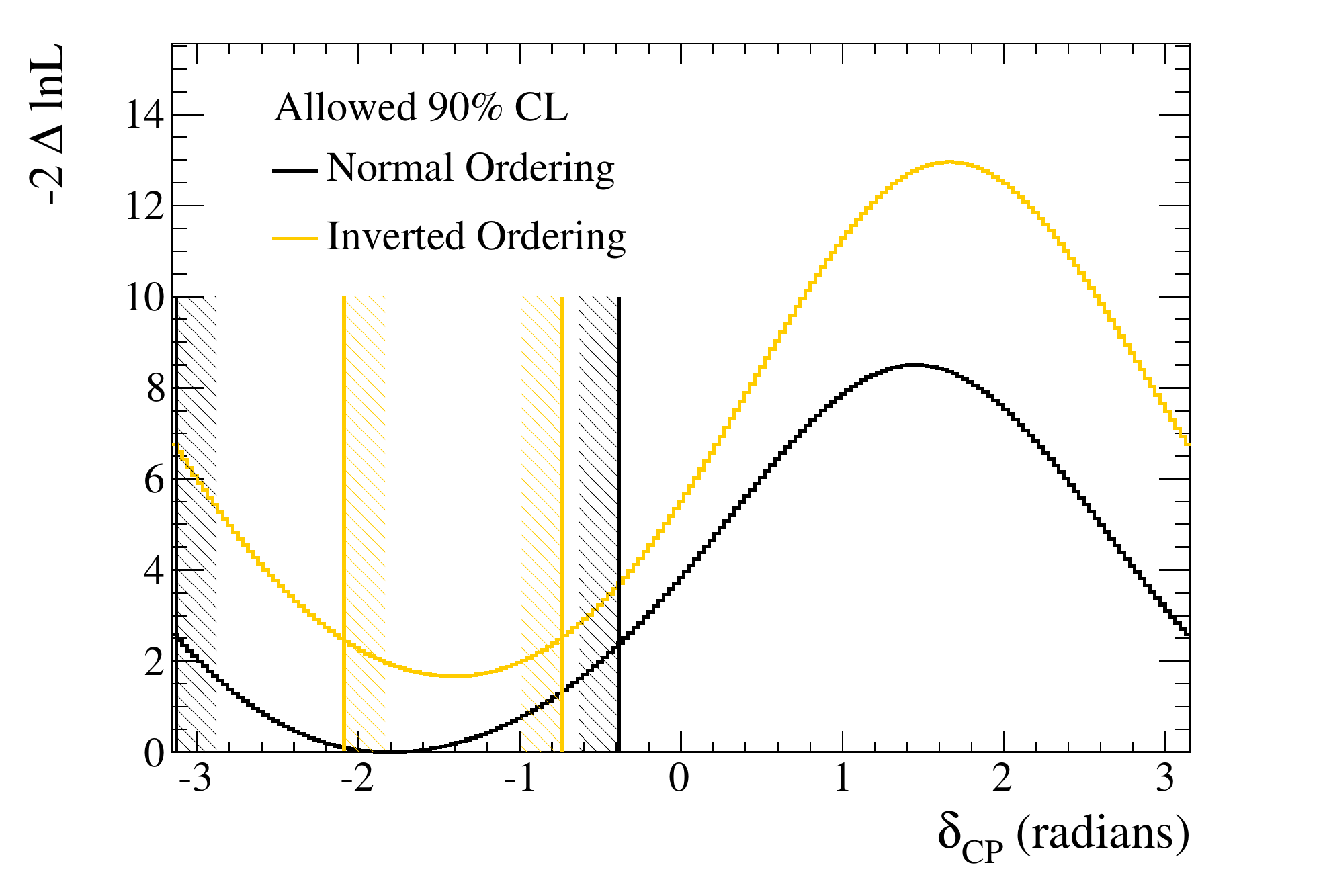}
\caption{
$-2 \Delta \ln \mathcal{L}$ as a function of $\delta_{CP}$ for the normal (black) and inverted (red) mass ordering. The vertical lines show the corresponding allowed 90\% confidence intervals, calculated using the Feldman-Cousins method. 
$\sin^2 \theta_{13}$ is marginalised using the reactor measurement as prior probability.}
\label{fig:fcdelta}
\end{figure}

Confidence intervals for $\delta_{CP}$ 
are obtained using the Feldman-Cousins method~\cite{Feldman:1997qc}. 
The parameter $\sin^2 \theta_{13}$ is marginalised using the reactor measurement.
The best-fit value is obtained for the normal ordering and $\delta_{CP} = -1.791$,
close to maximal CP violation (Fig.~\ref{fig:fcdelta}).
For inverted ordering the best-fit value of $\delta_{CP}$ is $-1.414$.
The hypothesis of CP conservation ($\delta_{CP}=0,\pi$) is excluded at 90\% C.L.
and $\delta_{CP} = 0$ is excluded at more than $2\sigma$.
The $\delta_{CP}$ confidence intervals at 90\% C.L. are ($-3.13$, $-0.39$) for normal ordering 
and ($-2.09$,$-0.74$) for inverted ordering. 
The Bayesian credible interval at 90\%, marginalising over the mass ordering,
is ($-3.13,-0.21$). The normal ordering is weakly favored 
over the inverted ordering with a posterior probability of 75\%.

Sensitivity studies show that, 
if the true value of $\delta_{CP}$ is $-\pi/2$ and the mass ordering 
is normal,
the fraction of pseudo-experiments where CP conservation ($\delta_{CP}=0,\pi$)
is excluded with a significance of 90\% C.L. 
is 17.3\%, with the amount of data used in this analysis.

\noindent \textit{Conclusions --- }
T2K has performed the first search for CP violation in neutrino
oscillations using $\nu_\mu \rightarrow \nu_e$ appearance and $\nu_\mu \rightarrow \nu_\mu$ disappearance channels in neutrino and antineutrino mode.
The one-dimensional confidence interval at 90\% for $\delta_{CP}$ spans the
range ($-3.13$, $-0.39$) in the normal mass ordering. The CP conservation hypothesis ($\delta_{CP}=0,\pi$)
is excluded at 90\% C.L. The data related to the measurements and results presented in this Letter can be found in Reference\cite{datarelease}.

We thank the J-PARC staff  for the superb accelerator performance and the CERN NA61/SHINE
collaboration  for  providing  valuable  particle  production  data.   We  acknowledge  the  support of MEXT, Japan; NSERC (grant SAPPJ-2014-00031), NRC and CFI, Canada; CEA
and CNRS/IN2P3, France; DFG, Germany; INFN, Italy; National Science Centre (NCN),
Poland; RSF, RFBR and MES, Russia; MINECO and ERDF funds, Spain; SNSF and SERI,
Switzerland;  STFC,  UK;  and  DOE,  USA.  We  also  thank  CERN  for  the  UA1/NOMAD
magnet, DESY for the HERA-B magnet mover system, NII for SINET4, the WestGrid and
SciNet consortia in Compute Canada, and GridPP, UK. In addition, participation of individual researchers and institutions has been further supported by funds from:  ERC (FP7),
H2020 RISE-GA644294-JENNIFER, EU; JSPS, Japan; Royal Society, UK; DOE Early Career program, USA.



%

\bibliography{biblio/reference}

\end{document}